\documentclass[compsoc,conference,a4paper,10pt,times]{IEEEtran}
\IEEEoverridecommandlockouts
\usepackage[numbers]{natbib}
\usepackage{amsmath,amssymb,amsfonts}
\usepackage{algorithmic}
\usepackage{graphicx}
\usepackage{textcomp}
\usepackage{bmpsize}
\usepackage{xcolor}
\usepackage{lipsum}
\PassOptionsToPackage{hyphens}{url}
\usepackage[colorlinks=true,urlcolor=black]{hyperref}
\def\BibTeX{{\rm B\kern-.05em{\sc i\kern-.025em b}\kern-.08em
    T\kern-.1667em\lower.7ex\hbox{E}\kern-.125emX}}

\hypersetup{breaklinks=true}

\usepackage{todonotes}

\title{Cybersecurity is the True Frontier for Generative AI Success or Failure }
\author{
\IEEEauthorblockN{
Edward Raff\IEEEauthorrefmark{1}\IEEEauthorrefmark{2},
Maor Ashkenazi\IEEEauthorrefmark{3},
Sagar Samtani\IEEEauthorrefmark{4},
David J. Elkind\IEEEauthorrefmark{1},
Sven Krasser\IEEEauthorrefmark{1}
}
\IEEEauthorblockA{\IEEEauthorrefmark{1}CrowdStrike}
\IEEEauthorblockA{\IEEEauthorrefmark{2}University of Maryland, Baltimore County}
\IEEEauthorblockA{\IEEEauthorrefmark{3}Department of Computer Science, Ben-Gurion University of the Negev}
\IEEEauthorblockA{\IEEEauthorrefmark{4}Data Science and Artificial Intelligence Lab, Kelley School of Business}
}

\begin{document}

\maketitle

\begin{abstract}
Cybersecurity is a real-life test-bed for many machine learning problems at once, especially when considering modern strides in using Large Language Models (LLMs) to automate processes as ``agents.'' Cybersecurity workflows require orchestrating hundreds of standard and bespoke tools through various formats. The scale of cybersecurity data is enormous; for example, a single malware sample can be viewed as a sequence of billions of tokens. The cost of labeling any file by experts is enormous and labor-intensive, in part because an adversary (possibly a well-funded nation state actor) is attempting to subvert your detection methods. Even skilled experts may disagree on the correct label, creating ambiguity in what constitutes ground truth. When deployed, models must run quickly on billions of items a day, where low-latency is critical for operational success, in a continuously changing environment. In addition, explainability is not optional: analysts demand clear reasoning for model decisions to cope with the large number of false-positive alerts they face daily, and to quickly develop remediation and understand how something went wrong. In short, the amount of complexity cybersecurity is greater than that of natural language and computer vision, and thus we posit that cybersecurity is the better test-case for general AI progress than other, well-studied fields. 
\end{abstract}

\section{Introduction}

Broadly, cybersecurity encompasses protecting a wide range of resources: on individual endpoints, in cloud resources, and across networks; in emails and document macros and in software; in PowerShell scripts and identity management. The challenges are daunting: evaluating security data is a highly specialized and challenging workflow, there is a shortage of domain experts \cite{isc2}, the volume of data (of which nearly all are completely unlabeled) continues to grow ~\cite{apruzzese2022sok}, and the threat environment evolves rapidly~\cite{ceschin2024machine,kshetri2025transforming}. 

Naturally, cybersecurity professionals have adopted automation strategies to help turn the tide. Artificial intelligence has made great advances in translation of both human and computer languages, code generation, summarizing large corpora of data, and retrieving relevant documents; however, it is not yet a panacea for cybersecurity defenders. And AI is a double-edged sword, as it can also provide attackers tools that enable them to probe defenses more systematically, generate and refine attacks at scale, and even target other AI or machine learning models themselves as part of the attack surface~\cite{guo2025frontieraisimpactcybersecurity,Vulpe2024}.

In this article, we argue that cybersecurity is a useful and fertile proving ground for the expansive claims of generative AI's proponents. As a motivating example, we will focus specifically on static analysis of software binaries. Static analysis is the task of determining whether a specific software executable is malicious. We examine four themes in which the application of current generative AI technologies to static analysis is, at best, incomplete. While generative AI holds great promise in these areas, it also comes with its own distinct challenges.

First, in \S \ref{sec:static}, we outline the static analysis task and the key factors which make it particularly challenging. 
Second, in \S \ref{sec:tool_use}, we will discuss the tool-using agents as an important avenue to real-world impact. 
Third, in \S \ref{sec:long_context} we will overview some of the purely technical roadblocks in applying transformer neural networks to software, with particular attention to long sequences.
Fourth, in \S \ref{sec:evaluation} we will discuss challenges in evaluation that prevent confident labeling and result in software corpora with a large amounts of unlabeled data. 

Finally, in \S \ref{sec:deployment}, we discuss how the continuously changing environments and tight constraints on human analysts create complex interactions with all the previously mentioned issues. However, this complex environment is also a true test-bed for self-adapting and learning systems to demonstrate real success. 

\section{Static Analysis Overview} \label{sec:static}

Static analysis, in general, is the task of inferring properties of code (source or compiled) without running the code itself. In short, it is an attempt to answer questions despite the halting problem preventing you from doing so in absolute terms. Static analysis of binary portable executable (PE) files is often used to determine whether an executable is malicious, similar to known prior samples, and to support other analyses that enable determining, preventing, and remediating the impacts of malicious files. Examples of malware include ransomware and remote access tools. In particular, static analysis of PE files involves inferring the program's operation from its contents on disk rather than from its behavior during execution. 
(By contrast, executing a piece of software and monitoring its actions is \textit{dynamic} analysis.) Although less than 20\% of attempted breaches used malware~\cite{gtr2026}, the static analysis task is straightforward to grasp and is well-studied in academic literature \cite{shi2013asc, rieck2008learning, dahl2013large, yin2007panorama, kolbitsch2009effective, perdisci2008mcboost, kolter2006learning, zhang2007metaaware, saxe2015deep}.

Neural networks and generative AI have demonstrated remarkable progress in fields such as computer vision and natural language. However, similar gains have not been realized for static analysis. We will briefly review some of the key factors that make static analysis particularly challenging as they are interconnected, before reviewing them again in more detail along the path of tool use through evaluation, and then deployment.

\textbf{Data size.} Entire datasets used in computer vision are smaller than a single software binary. While ImageNet is a standard benchmark dataset for computer vision, its 1,281,167 images are stored in only 17 GB after being resized to 256 pixels by 256 pixels. This widely-studied dataset is dwarfed by a single video game, which can easily exceed 40 GB in size. Thus, routine work in static analysis requires analyzing single samples that are an order of magnitude larger than entire \textit{datasets} from other fields. These analyses are not trivially parallelizable in the same way that many machine learning workloads are, further complicating operational factors. 

\textbf{Expertise.} Labeling malicious software requires understanding the myriad of different strategic objectives of malware authors, and a deep familiarity with the low-level tactics to achieve those objectives. These tactics include first-order things, such as the sequence of steps to achieve the goal, and second-order tactics, such as anti-analysis tricks and obfuscation to evade detection. In the industry, analysts are often specialized in a particular platform or domain where they have expertise, such as cloud security or macOS security, due to the ever-evolving, complex set of tools and attack vectors used across domains. 

\textbf{Time-consuming review.} A cybersecurity analyst with years of experience can take anywhere from hours to weeks of effort to analyze a single binary in detail (depending on the complexity of the binary and the analysis goals~\cite{Mohaisen:2013:UZA:2487788.2488056,Votipka2019}). By contrast, in the medical context, radiologists may spend a minute to up to 16 minutes to ``label'' a scan depending on the type of issue and experience~\cite{Sexauer2022,Forsberg2017-rt}. 

\textbf{Reverse engineering \& disassembly is hard.} Static analysis can involve reverse-engineering tasks such as disassembly (extracting the machine code from the compiled binary) and decompilation (analyzing the machine code to reconstruct the original source code). Neither task is as simple or straightforward as it might first appear. Utilities such as Linux's \texttt{objdump -d} employ a linear sweep algorithm to decode each byte sequentially; this process is error-prone for a number of reasons. The most straightforward reason is that the utility will attempt to decode bytes that are not machine code, such as embedded data or jump tables. Incorrect disassembly of even a single byte can cause the disassembler to become out of sync with the actual instructions. Reverse engineers use tools such as Ghidra, RADARE and IDA Pro, which recursively refine their proposed disassembly; even so, there is an art to this task because there can be ambiguities in decoding.

\subsection{Coding Assistants} \label{sec:coding_assistants}

Some may consider coding assistants to be significant progress and a representative path toward judging the greater success of generative AI. We argue that this is not the case. Though highly valuable, we argue that they do not pass the bar we seek. First, coding assistants operate in a non-static context: they run the actual code and have the original source to iterate on and improve the code itself. Second, the latter dynamic execution is possible in part because coding assistants run in a non-adversarial environment, where there is no overtly intentional actor attempting to deceive, hide, or cause negative effects when the individual code is run. Indeed, coding assistants are the easier task of translating the user's intent into a detailed specification (the code). Static analysis of potential malware involves taking the most difficult to parse form of specification, compiled machine code, and reverse engineering the full scope of intents and methods. 

Nor do coding assistants require extensive expert validation to understand if they are working correctly; a test suite can (partially) establish whether some code achieves the task. Coding is also less dynamic. Concept drift in coding assistants emerges slowly, as existing languages/libraries add new features or new languages enter widespread use. Finally, cybersecurity has unique requirements for zero-shot learning that coding assistants do not; a coding assistant does not have to dynamically re-assess how to solve a coding benchmark from one day to the next.

\section{Tool Use in Static Analysis} \label{sec:tool_use}

Tool-enabled AI is a promising research direction because tools are repeatable processes for solving a problem. Without tools, researchers are relying on the AI to correctly follow the correct sequence of operations each and every time it is invoked. Researchers can spend hours, or even months, of effort reverse engineering a \textit{single} binary executable, depending on the complexity of the file ~\cite{Votipka2019}. 
Even if automated systems are only as effective as a junior engineer (who may take up to three times longer than a senior reverse engineer to reverse engineer a piece of software~\cite{mantovani_re-mind_2022}), then the ability to scale agentic systems out over many machines would still be valuable in keeping up with cyber threats. 

A tool is purpose-built, usually deterministic, and should perform its function correctly. Tools can be command-line, graphical user interfaces, or web APIs. Tool use in cybersecurity presents a wide array of problems and opportunities, stemming from the diverse range of tools, their multi-modal interfaces, and a variety of one-off tools written by analysts. Tools have also been the focus of many significant open source and vendor-driven efforts, especially in an effort to help replicate workflows and produce systematic capabilities targeted around specific use cases. 

\subsection{Challenges to Tool Use}

Software decompilation and are examples of how tool use is a test-case for generative AI. Software decompilation decodes a software binary into a high-level language such as C++. In both cases, these are useful test-cases for generative AI tool use, because given the right setup a successfully decompiled binary can be \textit{recompiled} into the same binary, which is a challenge to make occur in practice~\cite{liu_superset_2026}, even before we consider obfuscated binaries~\cite{mohseni_can_2025,10.1145/2991079.2991114,Linn2003}. That said, there is some subtlety here: alternative compilers will compile the same code differently, and optimization levels and other settings on the compiler can change the resulting output~\cite{liu_assemblage_2024,saul_is_2024}. Despite this nuance, decompilation represents a useful north star for tool-using AI. Creating an AI training loop in which an AI system learns to operate a decompilation tool well enough that the decompiled code can be compiled into software with the \textit{same operation} as the original binary would represent a large step forward.

As natural language, computer code, and other cybersecurity data modalities can be represented as text, it is tempting to conflate the two domains. However, this similarity is superficial. As opposed to code, natural language admits a degree of ambiguity, but the intent can usually be inferred from context. Code is structured and has various semantics that must be obeyed to be valid; although LLMs have made recent strides on more complex structured generations~\cite{ugare_syncode_2024,firestone_utf-8_2025,geng_grammar-constrained_2023,scholak_picard_2021}, cybersecurity represents a new level of complexity. For cybersecurity, we are often concerned with the latent emergent structure of a system, rather than the intended structure of the designers. Coding errors, bugs, undocumented functionality, configuration settings, and even side-channel attacks create a ``shadow structure'' that is similar to, but distinct from, the one intended. In the static analysis setting, even elementary tasks like extracting metadata from the header of an executable can be maddeningly complicated. These complications arise because the operating systems often do not follow the published specification, and malware authors are attempting to exploit discrepancies and bugs~\cite{raff2017peheader}. Adversaries exploit these gaps between intention and software execution to evade detection or enable their malicious behavior. 

A second challenge lies in how the tools are incorporated into cybersecurity workflows. Ultimately, GenAI tools will be used by human cybersecurity professionals (even if the human users are at some remove), who will ultimately be responsible for their results. 

One key test for GenAI is whether the tool itself is sufficiently valuable or frictionless to be adopted. In a survey of 482 commonly used tools, ~\citet{mattei_qualitative_2022} noted that limitations in interaction support and usability hampered tool adoption. However, tool-using experts can achieve remarkable success, including perfectly decompiling small programs so that they can be recompiled exactly~\cite{burk_decomperson_2022}. 

Moreover, quantifying the productivity for GenAI tool use is challenging. In both cybersecurity and GenAI, it can be challenging to quantify the return on investment. In one study, cybersecurity professionals failed to benefit from GenAI or to understand when it made errors, and performed no better than learners without access to GenAI~\cite{mattei_im_2025}. Indeed, one of the major challenges with GenAI in cybersecurity is the high complexity and learning curve's impact on how novice users pick up and understand the tools used by a GenAI, as design patterns focusing on human evaluation can be subject to operator fatigue~\cite{raff_how_2025}. Prior studies on automation bias have found that even training humans on the risks of AI does not obviate their automation bias~\cite{suresh_misplaced_2020}. Firmly demonstrating that GenAI yields pronounced benefits for cybersecurity practitioners is a key step toward establishing its utility.

\subsection{Existing Promise}

Despite these challenges, there is promise in deploying AI tools in cybersecurity. Multiple cases have been presented in recent years, showing concrete improvements or at least potential for improving standard static analysis tools using GenAI, rather than fully supplanting current approaches. 

Cisco, the network technology company, published a study using multiple GenAI agents to analyze a relatively simple piece of malware~\cite{venere_using_2025}. Google, the internet advertising company, has shown similar, more preliminary public work on having LLMs operate in a fixed workflow to perform basic malware analysis tasks \cite{quintero}. While autonomous static analysis is a significant achievement, there are some limitations. The agents spent 46 minutes analyzing the single malware sample, which was not particularly large and did not contain significant countermeasures. That said, this application could still be a favorable tradeoff, when considering the cost relative to an expert performing the same task. An cybersecurity industry case study showed that even an imperfect tool can be valuable when professionals' working backlog is too large to complete all tasks within the required time frame~\cite{Raff2020autoyara}. Hybrid workflows in which triage systems direct simpler cases to AI and more challenging cases to humans could economically divide labor.

Well before LLMs rose to prominence, the United States' Defense Advanced Research Projects Agency (DARPA) hosted the Cyber Grand Challenge, a tournament designed to spur interest in and development of automated security systems. These systems resulted in the development of new and improved tooling ~\cite{nguyen-tuong_xandra_2018,shoshitaishvili_rise_2017,shoshitaishvili_mechanical_2018}, and though imperfect, provide strong evidence that automation is possible, and thus GenAI could be a valuable enhancement.
Another well-regarded tool is XBOW, a fully autonomous AI-driven penetration tester. Recently, XBOW achieved the \#1 on HackerRank\cite{xbow}, a community for bug-bounty programs. Though the methodology is not fully public, the results are a powerful demonstration of the power of autonomous tooling.

The recent history in the development of fuzzing tools also demonstrates the value of AI tooling. A fuzzing tool is used to uncover bugs by identifying specific inputs that can put a program into a broken or otherwise exploitable state. Conventional tools use classical AI search techniques \cite{Chen2018}. For over a decade, the ``American Fuzzy Lop'' (AFL) fuzzer has maintained its position as a state-of-the-art tool ~\cite{zalewski_american_2013,hazimeh_magma_2020,fioraldi_afl_2020}. This is an impressive achievement. However, AFL may be displaced if LLMs are able to more quickly discover plausible, buggy inputs ~\cite{chen_elfuzz_2025}, and it is plausible as LLMs can circumvent the issue fuzzers have in generating more complex input patterns to seed an analysis. 

Finally, existing static analysis tools themselves may be leveraged effectively by LLM-based agentic systems. As an example, the reverse engineering tool Ghidra was notable for providing production-level tooling for free, and Ghidra Model Context Protocol (MCP)\cite{ghidra_mcp} provides a means for both greater automation and lower user expertise to perform reverse engineering tasks.

\section{Long Context \& Non-Sequential Structure} \label{sec:long_context}

Cybersecurity is characterized by very large contexts. As previously discussed, static analysis a particularly acute example.\footnote{The problem of long context is not solely limited to static analysis. Other common examples of large data sizes in cybersecurity include phishing emails, vulnerability scan data, and firewall logs.} Tokenizing and analyzing all of the bytes in a sequence of $4 \cdot 10^{10}$ elements requires specialized tools or a truly gargantuan amount of RAM. While most executable files are not this large, it is nonetheless a regular occurrence on real networks and computers; indeed, one single datum in cybersecurity is typically larger than an entire dataset in another field. Even worse, malware authors are aware that large files can be challenging to analyze, and so padding a binary in various ways to make it larger and thus avoid analysis is a well-worn obfuscation strategy \cite{yuceel-obfuscate}. 

Many deep learning innovations, even if they have found success elsewhere, have been dashed upon the rocks of these cybersecurity challenges. Although publication bias has often concealed this fact, \citet{10.1145/3494110.3528242} has documented limitations of transformers on URLs and PE files. The case of URLs is especially interesting because, although they are relatively short, transformers still do not perform well, suggesting that the issue is not solely about sequence length. Similarly, ~\citet{MalConv,Raff2020b} showed that batch normalization degrades the performance of models trained in Windows PE files, even though batch norm was the \textit{de facto} standard for achieving \textit{deep} learning at publication time. 

Long context poses an acute dilemma for transformer-based architectures. Passing a large amount of context to the transformer can be computationally infeasible. This problem arises from the transformer's hallmark component, the self-attention mechanism. Self-attention requires pairwise comparisons of all of the input's tokens, which creates a memory requirement that grows quadratically with the size of the input. In turn, large memory requirements make it challenging to scale up transformer architectures to a single example, let alone a mini-batch~\cite{alam_holographic_2024,alam_recasting_2023,Wang2020,Choromanski2020}. On the other hand, truncating the long context can make the task computationally feasible, but deprives the model of crucial information.

Likewise, transformers are well-suited to natural language and computer vision tasks because both fields are characterized by their spatial correlations -- texts are read in order, and adjacent pixels in images often belong to the same object. In turn, this implies that limited context is needed because the most relevant information is nearby. LLMs, despite having large context windows, appear biased to using spatial context~\cite{hurwitz_large_2025}. But static analysis breaks this assumption because software contains jump instructions. Much like a \textit{Choose Your Own Adventure} book presents the reader with choices about which pages to read next, the branching logic of software can mean that two instructions are logically sequential, even if their corresponding bytes are not contiguous. (The reverse is also true: two instructions can be adjacent but not executed in the order of their bytes.)

~\citet{kunwar2025sok} provides an analysis of transformer-based approaches used for malware analysis, and provides evidence that integrating domain knowledge into transformers can be valuable. The most successful applications of transformers to cybersecurity tasks are instances in which researchers deviate from the standard vanilla transformer, modifying it with custom attention mechanisms and task-specific components~\cite{li2023iot, pi2023adatrans}. Other approaches consider adapting the transformer architecture to work directly on byte sequences \cite{florian, Horton2023BytesAA, pagnoni2024byte}. This also highlights an interesting counter-trend to the AI community at large, where transformers further removed inductive biases in favor of using more data. This strategy that may still work in cybersecurity, if the compute can be scaled to real-world data sizes. 

This issue of context length arises not only with processing raw bytes, but also when applying AI to generative tasks, such as disassembly.
Currently, state-of-the-art disassembly works by imposing structural constraints on bytes to match requirements of machine code~\cite{qin_tady_2025}. Auto-regressive generative AI fails to produce satisfying assembly on a whole-binary scale due to inconsistencies with the machine code specification~\cite{pei_xda_2021,yu_deepdi_2022}. And there is a long history of applying machine learning and optimization techniques to disassembly, such as search, linear programming, symbolic analysis, and more~\cite{flores-montoya_disassembly_2025}. There are tantalizing opportunities to construct an intelligent system that will make autonomous decisions about when and how to employ the different tools enumerated here. Merging deep learning with classical methods will hopefully improve results, reduce runtime, and also reduce the large amount of ``dark'' knowledge that goes into hand-tuning and maintaining these complex software systems for  disassembly and decompilation.

\section{Evaluation is Hard} \label{sec:evaluation}

Having now discussed the unique and readily available tools in cybersecurity, as well as the complexity of integrating cybersecurity data to effectively utilize these tools, we now turn to evaluating whether our labor has produced useful results. The overarching factor that makes static analysis cybersecurity hard is the difficulty of evaluation. The enormous effort demanded by static analysis is a reflection of just how challenging it is to infer the necessary information within this domain, and also why so much work goes into tooling~\cite{agtr,joyce_claravy_2025}.

Conventional machine learning approaches continue to empower very fruitful cybersecurity analysis. The popular TLSH~\cite{Oliver2013} digest function is used to identify files that are near duplicates. Index structures are used to improve search speeds over large malware corpora, and errors in this standard algorithm were found using LEAN; rewriting the acceleration structures sped this up even further~\cite{gonzalez_if_2025}. Machine learning algorithms are used in satisfiability module theories problems, as Boolean satisfiability problem solvers are used for various vulnerability analysis, exploit generation, and other tasks~\cite{vanegue_smt_2012,li_symbolic_2014,gupta_effective_2018}. These expensive methods are only feasible due to the high cost of having analysts perform this work, so GenAI is a natural area to explore for further benefits ~\cite{sun_smt_2023}.
Given how critical data labeling is to cybersecurity, we will elaborate more about the unique challenges that occur within this domain, and the unique concept drift issues that often make progress fleeting.

\subsection{Adversarially Induced Concept Drift}

Cybersecurity is inherently an adversarial domain: where adversaries iteratively improve and adapt their tactics to evade detection and achieve their goals. Simultaneously, as defenders are adopting powerful AI systems to automate workflows and provide new detection capabilities, adversaries are also conducting their own research and development to understand how new tools, techniques, and procedures can be used to avoid detection. This fact is often not highlighted in academic literature due to its applied nature and difficulty in measuring it. Adversaries refine and improve their methods each day, so creating a static snapshot of the world for a benchmark begets over-fitting to aspects of that snapshot of the world, rather than the attacker-defender dynamic. 

Naturally, there is also an adversarial dynamic in GenAI: malicious actors want to use these tools to further their aims, while the GenAI developers want to curtail certain activities. Thus, attackers have every incentive to produce ``out of distribution'' content that differs from what was in the training distribution, and thus subvert the GenAI guardrails. The adversarial dynamic also lends urgency to mitigating LLM ``jailbreaks.''\footnote{For instance, GenAI guardrails discourage the AI from outputting a recipe for a dangerous chemical. While this is completely understandable, the threat model posited for dangerous chemicals ignores ignores the practical limitations (skills, equipment, raw materials) and large security apparatus that exists to \textit{prevent} individuals from obtaining precursors~\cite{mehrotra_tree_2024, zhou_easyjailbreak_2024, pavlova_automated_2025}. In cybersecurity, there is no licensing requirement to operate computer, whereas there are licensing requirements for obtaining certain chemical compounds.} Additionally, common jailbreak obfuscations (base64) encoding are expected inputs within cybersecurity, reducing the utility of simple detection methods (detecting and removing base64 inputs). Moreover, conventional approaches, such as adversarial retraining of a classifier, can fundamentally change its decision boundary and cause mistakes on instances previously classified correctly. 

Moreover, while cybersecurity practitioners would like to leverage the promise of AI-based detectors to protect their customers, these AI systems may, themselves, provide a new attack surface. Similarly to how creating an AI agent to manage one's email correspondence raises the risk that an adversary could coax that same agent into sharing confidential materials over email, creating an AI cybersecurity agent that faithfully follows instructions while adapting to changing circumstances could be deceived or even suborned. Additionally, attackers can also find weaknesses in AI-based detectors and attack them directly \cite{skylight-cylance}. 

\subsection{Data \& Labels}\label{data-and-labels}

Adversarial concept drift likewise makes it hard to construct good training and evaluation sets, even within industry. First, concept drift creates strong temporal correlations~\cite{Joyce2021}, necessitating train-test splits in time to mitigate bias~\cite{235493,schvartzman2024new}. However, to stop \textit{current} threats, recent training data is significantly more useful than older data \cite{kantchelian2013approaches}. This observation requires an uncomfortable balance between exploiting temporal correlation to improve model performance, while avoiding deleterious drops in model performance once adversaries adapt and the distribution shifts. 

The adversarial and iterative nature of cybersecurity makes it challenging to collect timely and relevant cybersecurity data for modern attacks. New defenses compel attackers to develop new tactics, but are not the only source of concept drift. The inevitability of lapses in new software and systems provides an ever-changing landscape of computer security vulnerabilities. Even the incremental versions of operating systems means that similar software source code can yield vastly different binaries, further complicating efforts to generalize defenses. Managing information technology at even a small organization involves triaging software vulnerabilities, rolling out patches, and even \textit{delaying} patches to assure system stability. This also means that, from an LLM's perspective, knowledge about what systems are (not)  protected is not fixed in time, because those facts will change --- resulting in conflicting signals between the LLM's training data and any updated information. In-context learning, continuous pre-training, and other strategies may be viable mitigation methods, but need significant exploration and validation. 

Data accessibility to researchers outside of industry is also notoriously difficult. Whereas natural language and computer vision tasks have standard, widely available benchmark datasets, such as Penn TreeBank or CIFAR-10, intellectual property rights can prevent data-sharing and make it challenging for cybersecurity defenders to compile useful datasets. In the case of static analysis, software copyrights prevent sharing a corpus of benign software, as that would constitute piracy. 
\footnote{While permissively licensed projects can be useful in some respects, as they have fewer legal encumbrances, they are not a panacea. Classifiers trained solely on such projects have little context for evaluating proprietary software, creating a distribution shift when the classifier is applied to real systems. Even worse, this distribution shift risks errors on some of the most widely-used software in the world, which is not an acceptable outcome for an industrial cybersecurity company.}
While the \textit{legal} risk to sharing malware is minimal~\cite{joyce_ember2024_2025,Harang2020}, it creates risk for accidental execution. In other cases, cybersecurity data is encumbered due to its sensitive nature \cite{Krasser2025malware}. Moreover, because the most up-to-date detection mechanisms are part of a company's competitive advantage, companies tend to share only older or partial detection logic through platforms like VirusTotal, which further complicates academic researchers' ability to gather accurate ground-truth labels for model evaluation and training.

Rights holders are actively contesting whether GenAI outputs are intellectual property, or if GenAI itself can infringe on intellectual property rights, and this also has consequences for cybersecurity. GenAI licensing agreements often include provisions disallowing the GenAI outputs from being used to train another GenAI system, which will further complicate incorporating GenAI systems into workflows.  The ability for LLMs to generate potentially representative synthetic data or otherwise enable federated learning is an open question within cybersecurity. 

Finally, data sovereignty and data localization requirements preclude the construction of a centralized, globally representative dataset. Yet malware is frequently a tool of nation-state adversaries~\cite{Langner2011}, making data localization a further confounding factor for both industry and academia.

\section{Deployment Considerations} \label{sec:deployment}

Once one has gone through the effort to determine if a static analysis model is useful to deploy, so too do we encounter unique considerations and factors for deployment itself. There are numerous issues from real-time systems requirements, low overhead~\cite{Nguyen2021}, and others --- but we will focus on two in particular: few-shot learning and explainability. 

\subsection{Zero- \& Few-Shot Learning}
Unlike other domains, where the objects of study remain relatively consistent over short time spans, cybersecurity faces a constant stream of new malware and attack methods. Traditional supervised models quickly lose effectiveness in detecting cyberattack, so few-shot and zero-shot learning are important for detecting threats quickly~\cite{barros2022malware, akdeniz2025voltron}. This is not an artifact of just the difficulty in data acquisition, but also a requirement of any system, even when data is available. A customer's network will change day by day as users are added, systems are changed, and a myriad of regular and required operational activities perturb the network in large and small iterative ways. New compiler versions, build tools, and other software tools will change the data distribution in static analysis. All models will, in a relatively short time, be subject to a new out-of-distribution environment that does not closely resemble the original training distribution, even if the labeled data were perfect at the time of model training. 

As we mentioned, the latency for data labeling is measured in hours at best, so it becomes necessary to also consider unsupervised methods for adaptation. In particular, we are interested in unsupervised or semi-supervised methods for adapting models currently under deployment. For instance, \citet{zeng2024unleashing} demonstrates how contrastive learning can model network traffic and subsequently be applied to identify unusual patterns that signal potential attacks. The work of ~\citet{zeng2024unleashing} is also a pertinent reminder of the difficulties in data acquisition, as they worked with local utilities to obtain data. It is common for simulated data to also be used, but this imposes its own complexities in ensuring an accurate simulation --- as many such datasets have been criticized for lack of specificity to real-world cybersecurity at their time of development, let alone years or even decades later~\cite{siddique_kdd_2019,divekar_benchmarking_2018}.

\subsection {Explainable AI}
Explainability is critically important in cybersecurity because in these scenarios, AI systems operate on the front lines of defense, generating decisions and alerts that may impact the safety and operational continuity of organizations. An analyst working in a Security Operation Center (SOC) will experience hundreds or thousands of alerts producing ``alert fatigue'' ~\cite{gelman_that_2023,Raff2020d}. Being able to explain why an alert is likely correct, reducing the analyst's time to confirm, is of extreme value. Beyond confirmation, such explainability is also important for \textit{remediation}. Knowing how a deleterious event or actor came to infect a system, or gain movement through a network, is itself needed information to block re-entry, create patches, and other activities needed to improve security after the fact. 

Various works have been done in explainable AI for cybersecurity, and we expect more to come. \citet{sarker2024explainable} provides an in-depth review of explainable AI methods, applied to cybersecurity tasks from intrusion detection to malware analysis. The authors detail the challenges of high dimensionality, real-time constraints, adversarial threats, and balancing interpretability with detection performance. \cite{rastogi2025survey, capuano2022explainable} surveys the necessity of explainable AI for threat intelligence, anomaly detection, malware analysis and more. \cite{madamidola2025detecting, paltun2025robust, mohale2025evaluating} propose interpretable models for malware detection and intrusion detection. Others have turned to Graph Neural Networks (GNNs) to better represent the underlying data (e.g., a computer network is most naturally represented as a graph) and thus also obtain more interpretable explanations in cybersecurity contexts \cite{he2022illuminati}. GNNs have been powerful for examining some of the social aspects of cybersecurity, including cybercrime analysis, Dark Web-based cyber threat intelligence, vulnerability propagation in open source software platforms, and more. These contexts often lack significant labeled training data, and therefore require unsupervised and/or self-supervised learning paradigms.

\section{Conclusion} \label{sec:conclusion}

Cybersecurity is so complex that we have not yet covered all the issues within the sub-domain of static analysis that make it a challenge. Indeed, for an individual to independently perform work at this intersection encompasses an unreasonably broad spectrum of skills, ranging from low-level computer science, including operating systems, assembly, and networking, to high-level mathematics and algorithms. The cost of performing work is high due to the size and scale of the data, as well as additional legal barriers that must be respected. It is a domain that would likely be avoided if the impact and relevance were not so great: insufficient cybersecurity places private data at risk, ransomware funds adversarial regimes and criminal organizations, and has become the target of war due to targeting critical infrastructure like power plants. It is thus a space where more effort is warranted, with reward in having a positive impact on personal and national security, and a scientific playground to truly judge how far AI has come in the modern era. 

\bibliographystyle{IEEEtranN}
\bibliography{aaai2026}

\begin{thebibliography}{104}
\providecommand{\natexlab}[1]{#1}
\providecommand{\url}[1]{#1}
\csname url@samestyle\endcsname
\providecommand{\newblock}{\relax}
\providecommand{\bibinfo}[2]{#2}
\providecommand{\BIBentrySTDinterwordspacing}{\spaceskip=0pt\relax}
\providecommand{\BIBentryALTinterwordstretchfactor}{4}
\providecommand{\BIBentryALTinterwordspacing}{\spaceskip=\fontdimen2\font plus
\BIBentryALTinterwordstretchfactor\fontdimen3\font minus \fontdimen4\font\relax}
\providecommand{\BIBforeignlanguage}[2]{{%
\expandafter\ifx\csname l@#1\endcsname\relax
\typeout{** WARNING: IEEEtranN.bst: No hyphenation pattern has been}%
\typeout{** loaded for the language `#1'. Using the pattern for}%
\typeout{** the default language instead.}%
\else
\language=\csname l@#1\endcsname
\fi
#2}}
\providecommand{\BIBdecl}{\relax}
\BIBdecl

\bibitem[{International Information System Security Certification Consortium}(2024)]{isc2}
\BIBentryALTinterwordspacing
{International Information System Security Certification Consortium}. (2024, 09) Employers must act as cybersecurity workforce growth stalls and skills gaps widen employers must act as cybersecurity workforce growth stalls and skills gaps widen. [Online]. Available: \url{https://www.isc2.org/Insights/2024/09/Employers-Must-Act-Cybersecurity-Workforce-Growth-Stalls-as-Skills-Gaps-Widen}
\BIBentrySTDinterwordspacing

\bibitem[Apruzzese et~al.(2022)Apruzzese, Laskov, and Tastemirova]{apruzzese2022sok}
G.~Apruzzese, P.~Laskov, and A.~Tastemirova, ``Sok: The impact of unlabelled data in cyberthreat detection,'' in \emph{2022 IEEE 7th European Symposium on Security and Privacy (EuroS\&P)}.\hskip 1em plus 0.5em minus 0.4em\relax IEEE, 2022, pp. 20--42.

\bibitem[Ceschin et~al.(2024)Ceschin, Botacin, Bifet, Pfahringer, Oliveira, Gomes, and Gr{\'e}gio]{ceschin2024machine}
F.~Ceschin, M.~Botacin, A.~Bifet, B.~Pfahringer, L.~S. Oliveira, H.~M. Gomes, and A.~Gr{\'e}gio, ``Machine learning (in) security: A stream of problems,'' \emph{Digital Threats: Research and Practice}, vol.~5, no.~1, pp. 1--32, 2024.

\bibitem[Kshetri(2025)]{kshetri2025transforming}
N.~Kshetri, ``Transforming cybersecurity with agentic ai to combat emerging cyber threats,'' \emph{Telecommunications Policy}, p. 102976, 2025.

\bibitem[Guo et~al.(2025)Guo, Potter, Shi, Wang, Zhang, and Song]{guo2025frontieraisimpactcybersecurity}
\BIBentryALTinterwordspacing
W.~Guo, Y.~Potter, T.~Shi, Z.~Wang, A.~Zhang, and D.~Song, ``Frontier ai's impact on the cybersecurity landscape,'' 2025. [Online]. Available: \url{https://arxiv.org/abs/2504.05408}
\BIBentrySTDinterwordspacing

\bibitem[Vulpe et~al.(2024)Vulpe, Rughiniș, Țurcanu, and Rosner]{Vulpe2024}
\BIBentryALTinterwordspacing
S.-N. Vulpe, R.~Rughiniș, D.~Țurcanu, and D.~Rosner, ``Ai and cybersecurity: a risk society perspective,'' \emph{Frontiers in Computer Science}, vol.~6, Oct. 2024. [Online]. Available: \url{http://dx.doi.org/10.3389/fcomp.2024.1462250}
\BIBentrySTDinterwordspacing

\bibitem[CrowdStrike(2026)]{gtr2026}
\BIBentryALTinterwordspacing
CrowdStrike, ``2026 crowdstrike global threat report: Ai accelerates adversaries and reshapes the attack surface,'' CrowdStrike, Tech. Rep., 02 2026. [Online]. Available: \url{https://www.crowdstrike.com/en-us/global-threat-report/}
\BIBentrySTDinterwordspacing

\bibitem[Shi et~al.(2013)Shi, Que, Zhong, Meyer, Crenshaw, and He]{shi2013asc}
L.~Shi, J.~Que, Z.~Zhong, B.~Meyer, P.~Crenshaw, and Y.~He, ``A scalable implementation of malware detection based on network connection behaviors,'' in \emph{Proceedings of the International Conference on Cyber-Enabled Distributed Computing and Knowledge Discovery (CyberC)}, Oct 2013, pp. 59--66.

\bibitem[Rieck et~al.(2008)Rieck, Holz, Willems, D{\"u}ssel, and Laskov]{rieck2008learning}
K.~Rieck, T.~Holz, C.~Willems, P.~D{\"u}ssel, and P.~Laskov, ``Learning and classification of malware behavior,'' in \emph{Detection of Intrusions and Malware, and Vulnerability Assessment}.\hskip 1em plus 0.5em minus 0.4em\relax Springer, 2008, pp. 108--125.

\bibitem[Dahl et~al.(2013)Dahl, Stokes, Deng, and Yu]{dahl2013large}
G.~E. Dahl, J.~W. Stokes, L.~Deng, and D.~Yu, ``Large-scale malware classification using random projections and neural networks,'' in \emph{Acoustics, Speech and Signal Processing (ICASSP), 2013 IEEE International Conference on}.\hskip 1em plus 0.5em minus 0.4em\relax IEEE, 2013, pp. 3422--3426.

\bibitem[Yin et~al.(2007)Yin, Song, Egele, Kruegel, and Kirda]{yin2007panorama}
H.~Yin, D.~Song, M.~Egele, C.~Kruegel, and E.~Kirda, ``Panorama: capturing system-wide information flow for malware detection and analysis,'' in \emph{Proceedings of the 14th ACM conference on Computer and communications security}.\hskip 1em plus 0.5em minus 0.4em\relax ACM, 2007, pp. 116--127.

\bibitem[Kolbitsch et~al.(2009)Kolbitsch, Comparetti, Kruegel, Kirda, Zhou, and Wang]{kolbitsch2009effective}
C.~Kolbitsch, P.~M. Comparetti, C.~Kruegel, E.~Kirda, X.-y. Zhou, and X.~Wang, ``Effective and efficient malware detection at the end host.'' in \emph{USENIX security symposium}, 2009, pp. 351--366.

\bibitem[Perdisci et~al.(2008)Perdisci, Lanzi, and Lee]{perdisci2008mcboost}
R.~Perdisci, A.~Lanzi, and W.~Lee, ``Mcboost: Boosting scalability in malware collection and analysis using statistical classification of executables,'' in \emph{Computer Security Applications Conference, 2008. ACSAC 2008. Annual}.\hskip 1em plus 0.5em minus 0.4em\relax IEEE, 2008, pp. 301--310.

\bibitem[Kolter and Maloof(2006)]{kolter2006learning}
J.~Z. Kolter and M.~A. Maloof, ``Learning to detect and classify malicious executables in the wild,'' \emph{The Journal of Machine Learning Research}, vol.~7, pp. 2721--2744, 2006.

\bibitem[Zhang and Reeves(2007)]{zhang2007metaaware}
Q.~Zhang and D.~S. Reeves, ``Meta{A}ware: Identifying metamorphic malware,'' in \emph{Computer Security Applications Conference, 2007. ACSAC 2007. Twenty-Third Annual}.\hskip 1em plus 0.5em minus 0.4em\relax IEEE, 2007, pp. 411--420.

\bibitem[Saxe and Berlin(2015)]{saxe2015deep}
J.~Saxe and K.~Berlin, ``Deep neural network based malware detection using two dimensional binary program features,'' in \emph{2015 10th International Conference on Malicious and Unwanted Software (MALWARE)}, 2015, pp. 11--20.

\bibitem[Mohaisen and Alrawi(2013)]{Mohaisen:2013:UZA:2487788.2488056}
\BIBentryALTinterwordspacing
A.~Mohaisen and O.~Alrawi, ``Unveiling {Zeus}: {Automated} {Classification} of {Malware} {Samples},'' in \emph{Proceedings of the {22Nd} {International} {Conference} on {World} {Wide} {Web}}.\hskip 1em plus 0.5em minus 0.4em\relax New York, NY, USA: ACM, 2013, pp. 829--832, series Title: WWW '13 Companion. [Online]. Available: \url{http://doi.acm.org/10.1145/2487788.2488056}
\BIBentrySTDinterwordspacing

\bibitem[Votipka et~al.(2019)Votipka, Rabin, Micinski, Foster, and Mazurek]{Votipka2019}
D.~Votipka, S.~M. Rabin, K.~Micinski, J.~S. Foster, and M.~M. Mazurek, ``An {Observational} {Investigation} of {Reverse} {Engineers}' {Processes},'' in \emph{{USENIX} {Security} {Symposium}}, 2019.

\bibitem[Sexauer and Bestler(2022)]{Sexauer2022}
\BIBentryALTinterwordspacing
R.~Sexauer and C.~Bestler, ``Time is money: Considerations for measuring the radiological reading time,'' \emph{Journal of Imaging}, vol.~8, no.~8, p. 208, Jul. 2022. [Online]. Available: \url{http://dx.doi.org/10.3390/jimaging8080208}
\BIBentrySTDinterwordspacing

\bibitem[Forsberg et~al.(2017)Forsberg, Rosipko, and Sunshine]{Forsberg2017-rt}
D.~Forsberg, B.~Rosipko, and J.~L. Sunshine, ``\BIBforeignlanguage{en}{Radiologists' variation of time to read across different procedure types},'' \emph{\BIBforeignlanguage{en}{J. Digit. Imaging}}, vol.~30, no.~1, pp. 86--94, Feb. 2017.

\bibitem[Mantovani et~al.(2022)Mantovani, Aonzo, Fratantonio, and Balzarotti]{mantovani_re-mind_2022}
\BIBentryALTinterwordspacing
A.~Mantovani, S.~Aonzo, Y.~Fratantonio, and D.~Balzarotti, ``\BIBforeignlanguage{en}{\{{RE}-{Mind}\}: a {First} {Look} {Inside} the {Mind} of a {Reverse} {Engineer}},'' 2022, pp. 2727--2745. [Online]. Available: \url{https://www.usenix.org/conference/usenixsecurity22/presentation/mantovani}
\BIBentrySTDinterwordspacing

\bibitem[Liu et~al.(2026)Liu, Sun, Gilray, and Micinski]{liu_superset_2026}
\BIBentryALTinterwordspacing
C.~Liu, Y.~Sun, T.~Gilray, and K.~Micinski, ``Superset {Decompilation},'' Mar. 2026, arXiv:2603.28002 [cs]. [Online]. Available: \url{http://arxiv.org/abs/2603.28002}
\BIBentrySTDinterwordspacing

\bibitem[Mohseni et~al.(2025)Mohseni, Mohammadi, Tilwani, Saxena, Ndawula, Vema, Raff, and Gaur]{mohseni_can_2025}
\BIBentryALTinterwordspacing
S.~Mohseni, S.~Mohammadi, D.~Tilwani, Y.~Saxena, G.~K. Ndawula, S.~Vema, E.~Raff, and M.~Gaur, ``\BIBforeignlanguage{en}{Can {LLMs} {Obfuscate} {Code}? {A} {Systematic} {Analysis} of {Large} {Language} {Models} into {Assembly} {Code} {Obfuscation}},'' \emph{\BIBforeignlanguage{en}{Proceedings of the AAAI Conference on Artificial Intelligence}}, vol.~39, no.~23, pp. 24\,893--24\,901, Apr. 2025, number: 23. [Online]. Available: \url{https://ojs.aaai.org/index.php/AAAI/article/view/34672}
\BIBentrySTDinterwordspacing

\bibitem[Banescu et~al.(2016)Banescu, Collberg, Ganesh, Newsham, and Pretschner]{10.1145/2991079.2991114}
\BIBentryALTinterwordspacing
S.~Banescu, C.~Collberg, V.~Ganesh, Z.~Newsham, and A.~Pretschner, ``Code {Obfuscation} against {Symbolic} {Execution} {Attacks},'' in \emph{Proceedings of the 32nd {Annual} {Conference} on {Computer} {Security} {Applications}}.\hskip 1em plus 0.5em minus 0.4em\relax New York, NY, USA: Association for Computing Machinery, 2016, pp. 189--200, series Title: ACSAC ’16. [Online]. Available: \url{https://doi.org/10.1145/2991079.2991114}
\BIBentrySTDinterwordspacing

\bibitem[Linn and Debray(2003)]{Linn2003}
\BIBentryALTinterwordspacing
C.~Linn and S.~Debray, ``Obfuscation of executable code to improve resistance to static disassembly,'' in \emph{Proceedings of the 10th {ACM} conference on {Computer} and communication security - {CCS} '03}.\hskip 1em plus 0.5em minus 0.4em\relax New York, New York, USA: ACM Press, 2003, p. 290. [Online]. Available: \url{http://portal.acm.org/citation.cfm?doid=948109.948149}
\BIBentrySTDinterwordspacing

\bibitem[Liu et~al.(2024)Liu, Saul, Sun, Raff, Fuchs, Southard~Pantano, Holt, and Micinski]{liu_assemblage_2024}
\BIBentryALTinterwordspacing
C.~Liu, R.~Saul, Y.~Sun, E.~Raff, M.~Fuchs, T.~Southard~Pantano, J.~Holt, and K.~Micinski, ``\BIBforeignlanguage{en}{Assemblage: {Automatic} {Binary} {Dataset} {Construction} for {Machine} {Learning}},'' \emph{\BIBforeignlanguage{en}{Advances in Neural Information Processing Systems}}, vol.~37, pp. 58\,698--58\,715, Dec. 2024. [Online]. Available: \url{https://proceedings.neurips.cc/paper_files/paper/2024/hash/6bbefc73a187dd42e0dc065b4e7a0615-Abstract-Datasets_and_Benchmarks_Track.html}
\BIBentrySTDinterwordspacing

\bibitem[Saul et~al.(2024)Saul, Liu, Fleischmann, Zak, Micinski, Raff, and Holt]{saul_is_2024}
\BIBentryALTinterwordspacing
R.~Saul, C.~Liu, N.~Fleischmann, R.~Zak, K.~Micinski, E.~Raff, and J.~Holt, ``\BIBforeignlanguage{en}{Is {Function} {Similarity} {Over}-{Engineered}? {Building} a {Benchmark}},'' \emph{\BIBforeignlanguage{en}{Advances in Neural Information Processing Systems}}, vol.~37, pp. 21\,636--21\,655, Dec. 2024. [Online]. Available: \url{https://proceedings.neurips.cc/paper_files/paper/2024/hash/2663c994c84a79b338bca613fe1ae223-Abstract-Datasets_and_Benchmarks_Track.html}
\BIBentrySTDinterwordspacing

\bibitem[Ugare et~al.(2024)Ugare, Suresh, Kang, Misailovic, and Singh]{ugare_syncode_2024}
\BIBentryALTinterwordspacing
S.~Ugare, T.~Suresh, H.~Kang, S.~Misailovic, and G.~Singh, ``\BIBforeignlanguage{en}{{SynCode}: {LLM} {Generation} with {Grammar} {Augmentation}},'' \emph{\BIBforeignlanguage{en}{Transactions on Machine Learning Research}}, Nov. 2024. [Online]. Available: \url{https://openreview.net/forum?id=HiUZtgAPoH}
\BIBentrySTDinterwordspacing

\bibitem[Firestone et~al.(2025)Firestone, Ugare, Singh, and Misailovic]{firestone_utf-8_2025}
\BIBentryALTinterwordspacing
P.~Firestone, S.~Ugare, G.~Singh, and S.~Misailovic, ``\BIBforeignlanguage{en}{{UTF}-8 {Plumbing}: {Byte}-level {Tokenizers} {Unavoidably} {Enable} {LLMs} to {Generate} {Ill}-formed {UTF}-8},'' Aug. 2025. [Online]. Available: \url{https://openreview.net/forum?id=8ExXncFpf6#discussion}
\BIBentrySTDinterwordspacing

\bibitem[Geng et~al.(2023)Geng, Josifoski, Peyrard, and West]{geng_grammar-constrained_2023}
\BIBentryALTinterwordspacing
S.~Geng, M.~Josifoski, M.~Peyrard, and R.~West, ``Grammar-{Constrained} {Decoding} for {Structured} {NLP} {Tasks} without {Finetuning},'' in \emph{Proceedings of the 2023 {Conference} on {Empirical} {Methods} in {Natural} {Language} {Processing}}, H.~Bouamor, J.~Pino, and K.~Bali, Eds.\hskip 1em plus 0.5em minus 0.4em\relax Singapore: Association for Computational Linguistics, Dec. 2023, pp. 10\,932--10\,952. [Online]. Available: \url{https://aclanthology.org/2023.emnlp-main.674/}
\BIBentrySTDinterwordspacing

\bibitem[Scholak et~al.(2021)Scholak, Schucher, and Bahdanau]{scholak_picard_2021}
\BIBentryALTinterwordspacing
T.~Scholak, N.~Schucher, and D.~Bahdanau, ``{PICARD}: {Parsing} {Incrementally} for {Constrained} {Auto}-{Regressive} {Decoding} from {Language} {Models},'' in \emph{Proceedings of the 2021 {Conference} on {Empirical} {Methods} in {Natural} {Language} {Processing}}, M.-F. Moens, X.~Huang, L.~Specia, and S.~W.-t. Yih, Eds.\hskip 1em plus 0.5em minus 0.4em\relax Online and Punta Cana, Dominican Republic: Association for Computational Linguistics, Nov. 2021, pp. 9895--9901. [Online]. Available: \url{https://aclanthology.org/2021.emnlp-main.779/}
\BIBentrySTDinterwordspacing

\bibitem[Raff et~al.(2017)Raff, Sylvester, and Nicholas]{raff2017peheader}
\BIBentryALTinterwordspacing
E.~Raff, J.~Sylvester, and C.~Nicholas, ``Learning the {PE} {Header}, {Malware} {Detection} with {Minimal} {Domain} {Knowledge},'' in \emph{Proceedings of the 10th {ACM} {Workshop} on {Artificial} {Intelligence} and {Security}}.\hskip 1em plus 0.5em minus 0.4em\relax New York, NY, USA: ACM, 2017, pp. 121--132, series Title: AISec '17. [Online]. Available: \url{http://doi.acm.org/10.1145/3128572.3140442}
\BIBentrySTDinterwordspacing

\bibitem[Mattei et~al.(2022)Mattei, McLaughlin, Katcher, and Votipka]{mattei_qualitative_2022}
\BIBentryALTinterwordspacing
J.~Mattei, M.~McLaughlin, S.~Katcher, and D.~Votipka, ``A {Qualitative} {Evaluation} of {Reverse} {Engineering} {Tool} {Usability},'' in \emph{Proceedings of the 38th {Annual} {Computer} {Security} {Applications} {Conference}}, ser. {ACSAC} '22.\hskip 1em plus 0.5em minus 0.4em\relax New York, NY, USA: Association for Computing Machinery, Dec. 2022, pp. 619--631. [Online]. Available: \url{https://dl.acm.org/doi/10.1145/3564625.3567993}
\BIBentrySTDinterwordspacing

\bibitem[Burk et~al.(2022)Burk, Pagani, Kruegel, and Vigna]{burk_decomperson_2022}
\BIBentryALTinterwordspacing
K.~Burk, F.~Pagani, C.~Kruegel, and G.~Vigna, ``\BIBforeignlanguage{en}{Decomperson: {How} {Humans} {Decompile} and {What} {We} {Can} {Learn} {From} {It}},'' 2022, pp. 2765--2782. [Online]. Available: \url{https://www.usenix.org/conference/usenixsecurity22/presentation/burk}
\BIBentrySTDinterwordspacing

\bibitem[Mattei et~al.(2025)Mattei, Pellegrini, Soto, Bohuk, and Votipka]{mattei_im_2025}
\BIBentryALTinterwordspacing
J.~Mattei, C.~Pellegrini, M.~Soto, M.~S. Bohuk, and D.~Votipka, ``\BIBforeignlanguage{en}{"{I}'m trying to learn{\ldots}and {I}'m shooting myself in the foot": {Beginners}' {Struggles} {When} {Solving} {Binary} {Exploitation} {Exercises}},'' 2025, pp. 2867--2886. [Online]. Available: \url{https://www.usenix.org/conference/usenixsecurity25/presentation/mattei}
\BIBentrySTDinterwordspacing

\bibitem[Raff et~al.(2025)Raff, Farris, and Biderman]{raff_how_2025}
E.~Raff, D.~Farris, and S.~Biderman, \emph{How {Large} {Language} {Models} {Work}}.\hskip 1em plus 0.5em minus 0.4em\relax Shelter Island: Manning, 2025.

\bibitem[Suresh et~al.(2020)Suresh, Lao, and Liccardi]{suresh_misplaced_2020}
\BIBentryALTinterwordspacing
H.~Suresh, N.~Lao, and I.~Liccardi, ``Misplaced {Trust}: {Measuring} the {Interference} of {Machine} {Learning} in {Human} {Decision}-{Making},'' in \emph{Proceedings of the 12th {ACM} {Conference} on {Web} {Science}}, ser. {WebSci} '20.\hskip 1em plus 0.5em minus 0.4em\relax New York, NY, USA: Association for Computing Machinery, Jul. 2020, pp. 315--324. [Online]. Available: \url{https://dl.acm.org/doi/10.1145/3394231.3397922}
\BIBentrySTDinterwordspacing

\bibitem[Venere(2025)]{venere_using_2025}
\BIBentryALTinterwordspacing
G.~Venere, ``\BIBforeignlanguage{en}{Using {LLMs} as a reverse engineering sidekick},'' Jul. 2025. [Online]. Available: \url{https://blog.talosintelligence.com/using-llm-as-a-reverse-engineering-sidekick/}
\BIBentrySTDinterwordspacing

\bibitem[Quintero(2024)]{quintero}
\BIBentryALTinterwordspacing
B.~Quintero. (2024, 04) From assistant to analyst: The power of gemini 1.5 pro for malware analysis. [Online]. Available: \url{https://cloud.google.com/blog/topics/threat-intelligence/gemini-for-malware-analysis}
\BIBentrySTDinterwordspacing

\bibitem[Raff et~al.(2020{\natexlab{a}})Raff, Zak, Munoz, Fleming, Anderson, Filar, Nicholas, and Holt]{Raff2020autoyara}
\BIBentryALTinterwordspacing
E.~Raff, R.~Zak, G.~L. Munoz, W.~Fleming, H.~S. Anderson, B.~Filar, C.~Nicholas, and J.~Holt, ``Automatic {YARA} {Rule} {Generation} {Using} {Biclustering},'' in \emph{13th {ACM} {Workshop} on {Artificial} {Intelligence} and {Security} ({AISec}'20)}, 2020, arXiv: 2009.03779. [Online]. Available: \url{http://arxiv.org/abs/2009.03779}
\BIBentrySTDinterwordspacing

\bibitem[Nguyen-Tuong et~al.(2018)Nguyen-Tuong, Melski, Davidson, Co, Hawkins, Hiser, Morris, Nguyen, and Rizzi]{nguyen-tuong_xandra_2018}
\BIBentryALTinterwordspacing
A.~Nguyen-Tuong, D.~Melski, J.~W. Davidson, M.~Co, W.~Hawkins, J.~D. Hiser, D.~Morris, D.~Nguyen, and E.~Rizzi, ``Xandra: {An} {Autonomous} {Cyber} {Battle} {System} for the {Cyber} {Grand} {Challenge},'' \emph{IEEE Security \& Privacy}, vol.~16, no.~2, pp. 42--51, Mar. 2018. [Online]. Available: \url{https://ieeexplore.ieee.org/document/8328984}
\BIBentrySTDinterwordspacing

\bibitem[Shoshitaishvili et~al.(2017)Shoshitaishvili, Weissbacher, Dresel, Salls, Wang, Kruegel, and Vigna]{shoshitaishvili_rise_2017}
\BIBentryALTinterwordspacing
Y.~Shoshitaishvili, M.~Weissbacher, L.~Dresel, C.~Salls, R.~Wang, C.~Kruegel, and G.~Vigna, ``Rise of the {HaCRS}: {Augmenting} {Autonomous} {Cyber} {Reasoning} {Systems} with {Human} {Assistance},'' in \emph{Proceedings of the 2017 {ACM} {SIGSAC} {Conference} on {Computer} and {Communications} {Security}}, ser. {CCS} '17.\hskip 1em plus 0.5em minus 0.4em\relax New York, NY, USA: Association for Computing Machinery, Oct. 2017, pp. 347--362. [Online]. Available: \url{https://dl.acm.org/doi/10.1145/3133956.3134105}
\BIBentrySTDinterwordspacing

\bibitem[Shoshitaishvili et~al.(2018)Shoshitaishvili, Bianchi, Borgolte, Cama, Corbetta, Disperati, Dutcher, Grosen, Grosen, Machiry, Salls, Stephens, Wang, and Vigna]{shoshitaishvili_mechanical_2018}
\BIBentryALTinterwordspacing
Y.~Shoshitaishvili, A.~Bianchi, K.~Borgolte, A.~Cama, J.~Corbetta, F.~Disperati, A.~Dutcher, J.~Grosen, P.~Grosen, A.~Machiry, C.~Salls, N.~Stephens, R.~Wang, and G.~Vigna, ``Mechanical {Phish}: {Resilient} {Autonomous} {Hacking},'' \emph{IEEE Security \& Privacy}, vol.~16, no.~2, pp. 12--22, Mar. 2018. [Online]. Available: \url{https://ieeexplore.ieee.org/abstract/document/8328966}
\BIBentrySTDinterwordspacing

\bibitem[Waisman(2025)]{xbow}
\BIBentryALTinterwordspacing
N.~Waisman. (2025, 06) The road to top 1: How xbow did it. [Online]. Available: \url{https://xbow.com/blog/top-1-how-xbow-did-it}
\BIBentrySTDinterwordspacing

\bibitem[Chen and Chen(2018)]{Chen2018}
P.~Chen and H.~Chen, ``2018 ieee symposium on security and privacy {(SP)},'' pp. 711--725, May 2018, arXiv: 1803.01307.

\bibitem[Zalewski(2013)]{zalewski_american_2013}
\BIBentryALTinterwordspacing
M.~Zalewski, ``american fuzzy lop,'' Nov. 2013. [Online]. Available: \url{https://lcamtuf.coredump.cx/afl/}
\BIBentrySTDinterwordspacing

\bibitem[Hazimeh et~al.(2020)Hazimeh, Herrera, and Payer]{hazimeh_magma_2020}
\BIBentryALTinterwordspacing
A.~Hazimeh, A.~Herrera, and M.~Payer, ``Magma: {A} {Ground}-{Truth} {Fuzzing} {Benchmark},'' \emph{Proc. ACM Meas. Anal. Comput. Syst.}, vol.~4, no.~3, pp. 49:1--49:29, Nov. 2020. [Online]. Available: \url{https://doi.org/10.1145/3428334}
\BIBentrySTDinterwordspacing

\bibitem[Fioraldi et~al.(2020)Fioraldi, Maier, Ei{\ss}feldt, and Heuse]{fioraldi_afl_2020}
\BIBentryALTinterwordspacing
A.~Fioraldi, D.~Maier, H.~Ei{\ss}feldt, and M.~Heuse, ``\BIBforeignlanguage{en}{{AFL}++ : {Combining} {Incremental} {Steps} of {Fuzzing} {Research}},'' 2020. [Online]. Available: \url{https://www.usenix.org/conference/woot20/presentation/fioraldi}
\BIBentrySTDinterwordspacing

\bibitem[Chen et~al.(2025)Chen, Dolan-Gavitt, and Lin]{chen_elfuzz_2025}
\BIBentryALTinterwordspacing
C.~Chen, B.~Dolan-Gavitt, and Z.~Lin, ``{ELFuzz}: Efficient input generation via {LLM-driven} synthesis over fuzzer space.'' in \emph{{34th USENIX} Security Symposium ({USENIX Security} 25)}, Jul. 2025, pp. 6279--6298. [Online]. Available: \url{http://arxiv.org/abs/2506.10323}
\BIBentrySTDinterwordspacing

\bibitem[Wired(2025)]{ghidra_mcp}
\BIBentryALTinterwordspacing
L.~Wired. (2025, 06) ghidramcp. [Online]. Available: \url{https://github.com/LaurieWired/GhidraMCP}
\BIBentrySTDinterwordspacing

\bibitem[Yuceel(April 09, 2024)]{yuceel-obfuscate}
\BIBentryALTinterwordspacing
H.~C. Yuceel. (April 09, 2024) The {MITRE ATT\&CK T1027} obfuscated files or information technique. [Online]. Available: \url{https://www.picussecurity.com/resource/the-mitre-attck-t1027-obfuscated-files-or-information-technique}
\BIBentrySTDinterwordspacing

\bibitem[Rudd et~al.(2022)Rudd, Rahman, and Tully]{10.1145/3494110.3528242}
\BIBentryALTinterwordspacing
E.~M. Rudd, M.~S. Rahman, and P.~Tully, ``Transformers for {End}-to-{End} {InfoSec} {Tasks}: {A} {Feasibility} {Study},'' in \emph{Proceedings of the 1st {Workshop} on {Robust} {Malware} {Analysis}}.\hskip 1em plus 0.5em minus 0.4em\relax New York, NY, USA: Association for Computing Machinery, 2022, pp. 21--31, series Title: WoRMA '22. [Online]. Available: \url{https://doi.org/10.1145/3494110.3528242}
\BIBentrySTDinterwordspacing

\bibitem[Raff et~al.(2018)Raff, Barker, Sylvester, Brandon, Catanzaro, and Nicholas]{MalConv}
\BIBentryALTinterwordspacing
E.~Raff, J.~Barker, J.~Sylvester, R.~Brandon, B.~Catanzaro, and C.~Nicholas, ``Malware {Detection} by {Eating} a {Whole} {EXE},'' in \emph{{AAAI} {Workshop} on {Artificial} {Intelligence} for {Cyber} {Security}}, Oct. 2018, arXiv: 1710.09435. [Online]. Available: \url{http://arxiv.org/abs/1710.09435}
\BIBentrySTDinterwordspacing

\bibitem[Raff et~al.(2021)Raff, Fleshman, Zak, Anderson, Filar, and McLean]{Raff2020b}
\BIBentryALTinterwordspacing
E.~Raff, W.~Fleshman, R.~Zak, H.~S. Anderson, B.~Filar, and M.~McLean, ``Classifying {Sequences} of {Extreme} {Length} with {Constant} {Memory} {Applied} to {Malware} {Detection},'' in \emph{The {Thirty}-{Fifth} {AAAI} {Conference} on {Artificial} {Intelligence}}, 2021, arXiv: 2012.09390. [Online]. Available: \url{http://arxiv.org/abs/2012.09390}
\BIBentrySTDinterwordspacing

\bibitem[Alam et~al.(2024)Alam, Raff, Biderman, Oates, and Holt]{alam_holographic_2024}
\BIBentryALTinterwordspacing
M.~M. Alam, E.~Raff, S.~R. Biderman, T.~Oates, and J.~Holt, ``\BIBforeignlanguage{en}{Holographic {Global} {Convolutional} {Networks} for {Long}-{Range} {Prediction} {Tasks} in {Malware} {Detection}},'' in \emph{\BIBforeignlanguage{en}{Proceedings of {The} 27th {International} {Conference} on {Artificial} {Intelligence} and {Statistics}}}.\hskip 1em plus 0.5em minus 0.4em\relax PMLR, Apr. 2024, pp. 4042--4050. [Online]. Available: \url{https://proceedings.mlr.press/v238/mahmudul-alam24a.html}
\BIBentrySTDinterwordspacing

\bibitem[Alam et~al.(2023)Alam, Raff, Biderman, Oates, and Holt]{alam_recasting_2023}
\BIBentryALTinterwordspacing
M.~M. Alam, E.~Raff, S.~Biderman, T.~Oates, and J.~Holt, ``\BIBforeignlanguage{en}{Recasting {Self}-{Attention} with {Holographic} {Reduced} {Representations}},'' in \emph{\BIBforeignlanguage{en}{Proceedings of the 40th {International} {Conference} on {Machine} {Learning}}}.\hskip 1em plus 0.5em minus 0.4em\relax PMLR, Jul. 2023, pp. 490--507. [Online]. Available: \url{https://proceedings.mlr.press/v202/alam23a.html}
\BIBentrySTDinterwordspacing

\bibitem[Wang et~al.(2020)Wang, Li, Khabsa, Fang, and Ma]{Wang2020}
\BIBentryALTinterwordspacing
S.~Wang, B.~Li, M.~Khabsa, H.~Fang, and H.~Ma, ``Linformer: {Self}-{Attention} with {Linear} {Complexity},'' vol. 2048, no. 2019, 2020, arXiv: 2006.04768. [Online]. Available: \url{http://arxiv.org/abs/2006.04768}
\BIBentrySTDinterwordspacing

\bibitem[Choromanski et~al.(2020)Choromanski, Likhosherstov, Dohan, Song, Gane, Sarlos, Hawkins, Davis, Mohiuddin, Kaiser, Belanger, Colwell, and Weller]{Choromanski2020}
\BIBentryALTinterwordspacing
K.~Choromanski, V.~Likhosherstov, D.~Dohan, X.~Song, A.~Gane, T.~Sarlos, P.~Hawkins, J.~Davis, A.~Mohiuddin, L.~Kaiser, D.~Belanger, L.~Colwell, and A.~Weller, ``Rethinking {Attention} with {Performers},'' pp. 1--38, 2020, arXiv: 2009.14794. [Online]. Available: \url{http://arxiv.org/abs/2009.14794}
\BIBentrySTDinterwordspacing

\bibitem[Hurwitz et~al.(2025)Hurwitz, Nicholas, and Raff]{hurwitz_large_2025}
\BIBentryALTinterwordspacing
J.~Hurwitz, C.~Nicholas, and E.~Raff, ``\BIBforeignlanguage{en}{Large {Language} {Models} and {Normalized} {Compression} {Distance}: {Better} {Compression} {Yet} {Worse} {Accuracy}},'' in \emph{\BIBforeignlanguage{en}{{ECAI} 2025}}.\hskip 1em plus 0.5em minus 0.4em\relax IOS Press, 2025, pp. 4273--4280. [Online]. Available: \url{https://ebooks.iospress.nl/doi/10.3233/FAIA251322}
\BIBentrySTDinterwordspacing

\bibitem[Kunwar et~al.(2025)Kunwar, Aryal, Gupta, Abdelsalam, and Bertino]{kunwar2025sok}
P.~Kunwar, K.~Aryal, M.~Gupta, M.~Abdelsalam, and E.~Bertino, ``Sok: Leveraging transformers for malware analysis,'' \emph{IEEE Transactions on Dependable and Secure Computing}, 2025.

\bibitem[Li and Li(2023)]{li2023iot}
Y.~Li and Y.~Li, ``Iot malware threat hunting method based on improved transformer,'' \emph{International Journal of Network Security}, vol.~25, no.~2, pp. 267--276, 2023.

\bibitem[Pi et~al.(2023)Pi, Tian, Pei, Chen, Wang, and Wang]{pi2023adatrans}
F.~Pi, S.~Tian, X.~Pei, P.~Chen, X.~Wang, and X.~Wang, ``Adatrans: An adaptive transformer for iot malware detection based on sensitive api call graph and inter-component communication analysis,'' \emph{Journal of Intelligent \& Fuzzy Systems}, vol.~45, no.~6, pp. 11\,439--11\,452, 2023.

\bibitem[St{\"o}rtz(2025)]{florian}
\BIBentryALTinterwordspacing
F.~St{\"o}rtz. (2025, 03) Byte back: Next-generation malware classification using binary transformers. [Online]. Available: \url{https://www.crowdstrike.com/en-us/blog/byte-back-next-gen-malware-classification/}
\BIBentrySTDinterwordspacing

\bibitem[Horton et~al.(2023)Horton, Mehta, Farhadi, and Rastegari]{Horton2023BytesAA}
M.~Horton, S.~Mehta, A.~Farhadi, and M.~Rastegari, ``Bytes are all you need: Transformers operating directly on file bytes,'' \emph{ArXiv}, vol. abs/2306.00238, 2023.

\bibitem[Pagnoni et~al.(2024)Pagnoni, Pasunuru, Rodriguez, Nguyen, Muller, Li, Zhou, Yu, Weston, Zettlemoyer, et~al.]{pagnoni2024byte}
A.~Pagnoni, R.~Pasunuru, P.~Rodriguez, J.~Nguyen, B.~Muller, M.~Li, C.~Zhou, L.~Yu, J.~Weston, L.~Zettlemoyer \emph{et~al.}, ``Byte latent transformer: Patches scale better than tokens,'' \emph{arXiv preprint arXiv:2412.09871}, 2024.

\bibitem[Qin et~al.(2025)Qin, Yang, Wang, Zhang, Gao, Zhang, and Chen]{qin_tady_2025}
\BIBentryALTinterwordspacing
S.~Qin, F.~Yang, H.~Wang, B.~Zhang, Z.~Gao, C.~Zhang, and K.~Chen, ``Tady: {A} {Neural} {Disassembler} without {Structural} {Constraint} {Violations},'' in \emph{{34th USENIX} Security Symposium ({USENIX Security} 25)}.\hskip 1em plus 0.5em minus 0.4em\relax arXiv, Jun. 2025, pp. 451--468, arXiv:2506.13323 [cs]. [Online]. Available: \url{http://arxiv.org/abs/2506.13323}
\BIBentrySTDinterwordspacing

\bibitem[Pei et~al.(2021)Pei, Guan, Williams-King, Yang, and Jana]{pei_xda_2021}
\BIBentryALTinterwordspacing
K.~Pei, J.~Guan, D.~Williams-King, J.~Yang, and S.~Jana, ``\BIBforeignlanguage{en}{{XDA}: {Accurate}, {Robust} {Disassembly} with {Transfer} {Learning}},'' in \emph{\BIBforeignlanguage{en}{Proceedings 2021 {Network} and {Distributed} {System} {Security} {Symposium}}}.\hskip 1em plus 0.5em minus 0.4em\relax Virtual: Internet Society, 2021. [Online]. Available: \url{https://www.ndss-symposium.org/wp-content/uploads/ndss2021_1B-3_23112_paper.pdf}
\BIBentrySTDinterwordspacing

\bibitem[Yu et~al.(2022)Yu, Qu, Hu, and Yin]{yu_deepdi_2022}
\BIBentryALTinterwordspacing
S.~Yu, Y.~Qu, X.~Hu, and H.~Yin, ``\BIBforeignlanguage{en}{{DeepDi}: {Learning} a {Relational} {Graph} {Convolutional} {Network} {Model} on {Instructions} for {Fast} and {Accurate} {Disassembly}},'' 2022, pp. 2709--2725. [Online]. Available: \url{https://www.usenix.org/conference/usenixsecurity22/presentation/yu-sheng}
\BIBentrySTDinterwordspacing

\bibitem[Flores-Montoya et~al.(2025)Flores-Montoya, Lim, Seitz, Sood, Raff, and Holt]{flores-montoya_disassembly_2025}
\BIBentryALTinterwordspacing
A.~Flores-Montoya, J.~Lim, A.~Seitz, A.~Sood, E.~Raff, and J.~Holt, ``Disassembly as {Weighted} {Interval} {Scheduling} with {Learned} {Weights},'' in \emph{2025 {IEEE} {Symposium} on {Security} and {Privacy} ({SP})}, May 2025, pp. 3033--3050, iSSN: 2375-1207. [Online]. Available: \url{https://ieeexplore.ieee.org/document/11023516}
\BIBentrySTDinterwordspacing

\bibitem[Joyce et~al.(2021{\natexlab{a}})Joyce, Raff, and Nicholas]{agtr}
R.~J. Joyce, E.~Raff, and C.~Nicholas, ``A {Framework} for {Cluster} and {Classifier} {Evaluation} in the {Absence} of {Reference} {Labels},'' in \emph{Proceedings of the 14th {ACM} {Workshop} on {Artificial} {Intelligence} and {Security} ({AISec} '21)}.\hskip 1em plus 0.5em minus 0.4em\relax Association for Computing Machinery, 2021, arXiv: 2109.11126v1.

\bibitem[Joyce et~al.(2025{\natexlab{a}})Joyce, Everett, Fuchs, Raff, and Holt]{joyce_claravy_2025}
\BIBentryALTinterwordspacing
R.~J. Joyce, D.~Everett, M.~Fuchs, E.~Raff, and J.~Holt, ``{ClarAVy}: {A} {Tool} for {Scalable} and {Accurate} {Malware} {Family} {Labeling},'' in \emph{Companion {Proceedings} of the {ACM} on {Web} {Conference} 2025}, ser. {WWW} '25.\hskip 1em plus 0.5em minus 0.4em\relax New York, NY, USA: Association for Computing Machinery, May 2025, pp. 277--286. [Online]. Available: \url{https://dl.acm.org/doi/10.1145/3701716.3715212}
\BIBentrySTDinterwordspacing

\bibitem[Oliver et~al.(2013)Oliver, Cheng, and Chen]{Oliver2013}
\BIBentryALTinterwordspacing
J.~Oliver, C.~Cheng, and Y.~Chen, ``{TLSH} -- {A} {Locality} {Sensitive} {Hash},'' in \emph{2013 {Fourth} {Cybercrime} and {Trustworthy} {Computing} {Workshop}}.\hskip 1em plus 0.5em minus 0.4em\relax IEEE, Nov. 2013, pp. 7--13. [Online]. Available: \url{http://ieeexplore.ieee.org/document/6754635/}
\BIBentrySTDinterwordspacing

\bibitem[Gonzalez(2025)]{gonzalez_if_2025}
\BIBentryALTinterwordspacing
J.~Gonzalez, ``If at first you don't succeed, trie, trie again: {Correcting} {TLSH} scalability claims for large-dataset malware forensics,'' \emph{Forensic Science International: Digital Investigation}, vol.~53, p. 301922, Jul. 2025. [Online]. Available: \url{https://www.sciencedirect.com/science/article/pii/S2666281725000617}
\BIBentrySTDinterwordspacing

\bibitem[Vanegue et~al.(2012)Vanegue, Heelan, and Rolles]{vanegue_smt_2012}
\BIBentryALTinterwordspacing
J.~Vanegue, S.~Heelan, and R.~Rolles, ``{SMT} solvers for software security,'' in \emph{Proceedings of the 6th {USENIX} conference on {Offensive} {Technologies}}, ser. {WOOT}'12.\hskip 1em plus 0.5em minus 0.4em\relax USA: USENIX Association, Aug. 2012, p.~9. [Online]. Available: \url{https://www.usenix.org/system/files/conference/woot12/woot12-final26.pdf}
\BIBentrySTDinterwordspacing

\bibitem[Li et~al.(2014)Li, Albarghouthi, Kincaid, Gurfinkel, and Chechik]{li_symbolic_2014}
\BIBentryALTinterwordspacing
Y.~Li, A.~Albarghouthi, Z.~Kincaid, A.~Gurfinkel, and M.~Chechik, ``Symbolic optimization with {SMT} solvers,'' \emph{SIGPLAN Not.}, vol.~49, no.~1, pp. 607--618, Jan. 2014. [Online]. Available: \url{https://dl.acm.org/doi/10.1145/2578855.2535857}
\BIBentrySTDinterwordspacing

\bibitem[Gupta et~al.(2018)Gupta, Saxena, Mahajan, and Bansal]{gupta_effective_2018}
S.~Gupta, A.~Saxena, A.~Mahajan, and S.~Bansal, ``\BIBforeignlanguage{en}{Effective {Use} of {SMT} {Solvers} for {Program} {Equivalence} {Checking} {Through} {Invariant}-{Sketching} and {Query}-{Decomposition}},'' in \emph{\BIBforeignlanguage{en}{Theory and {Applications} of {Satisfiability} {Testing} -- {SAT} 2018}}, O.~Beyersdorff and C.~M. Wintersteiger, Eds.\hskip 1em plus 0.5em minus 0.4em\relax Cham: Springer International Publishing, 2018, pp. 365--382.

\bibitem[Sun et~al.(2023)Sun, Yang, Wang, Wen, Jia, and Zhou]{sun_smt_2023}
\BIBentryALTinterwordspacing
M.~Sun, Y.~Yang, Y.~Wang, M.~Wen, H.~Jia, and Y.~Zhou, ``{SMT} {Solver} {Validation} {Empowered} by {Large} {Pre}-{Trained} {Language} {Models},'' in \emph{2023 38th {IEEE}/{ACM} {International} {Conference} on {Automated} {Software} {Engineering} ({ASE})}, Sep. 2023, pp. 1288--1300, iSSN: 2643-1572. [Online]. Available: \url{https://ieeexplore.ieee.org/document/10298442}
\BIBentrySTDinterwordspacing

\bibitem[Mehrotra et~al.(2024)Mehrotra, Zampetakis, Kassianik, Nelson, Anderson, Singer, and Karbasi]{mehrotra_tree_2024}
\BIBentryALTinterwordspacing
A.~Mehrotra, M.~Zampetakis, P.~Kassianik, B.~Nelson, H.~Anderson, Y.~Singer, and A.~Karbasi, ``\BIBforeignlanguage{en}{Tree of {Attacks}: {Jailbreaking} {Black}-{Box} {LLMs} {Automatically}},'' \emph{\BIBforeignlanguage{en}{Advances in Neural Information Processing Systems}}, vol.~37, pp. 61\,065--61\,105, Dec. 2024. [Online]. Available: \url{https://proceedings.neurips.cc/paper_files/paper/2024/hash/70702e8cbb4890b4a467b984ae59828a-Abstract-Conference.html}
\BIBentrySTDinterwordspacing

\bibitem[Zhou et~al.(2024)Zhou, Wang, Xiong, Xia, Gu, Chai, Zhu, Huang, Dou, Xi, Zheng, Gao, Zou, Yan, Le, Wang, Li, Shao, Gui, Zhang, and Huang]{zhou_easyjailbreak_2024}
\BIBentryALTinterwordspacing
W.~Zhou, X.~Wang, L.~Xiong, H.~Xia, Y.~Gu, M.~Chai, F.~Zhu, C.~Huang, S.~Dou, Z.~Xi, R.~Zheng, S.~Gao, Y.~Zou, H.~Yan, Y.~Le, R.~Wang, L.~Li, J.~Shao, T.~Gui, Q.~Zhang, and X.~Huang, ``{EasyJailbreak}: {A} {Unified} {Framework} for {Jailbreaking} {Large} {Language} {Models},'' Mar. 2024, arXiv:2403.12171 [cs]. [Online]. Available: \url{http://arxiv.org/abs/2403.12171}
\BIBentrySTDinterwordspacing

\bibitem[Pavlova et~al.(2025)Pavlova, Brinkman, Iyer, Albiero, Bitton, Nguyen, Ferrer, Evtimov, and Grattafiori]{pavlova_automated_2025}
\BIBentryALTinterwordspacing
M.~Pavlova, E.~Brinkman, K.~Iyer, V.~Albiero, J.~Bitton, H.~Nguyen, C.~C. Ferrer, I.~Evtimov, and A.~Grattafiori, ``\BIBforeignlanguage{en}{Automated {Red} {Teaming} with {GOAT}: the {Generative} {Offensive} {Agent} {Tester}},'' Jun. 2025. [Online]. Available: \url{https://openreview.net/forum?id=bDBnd9T2Cz}
\BIBentrySTDinterwordspacing

\bibitem[Ashkenazy and Zini(2019)]{skylight-cylance}
\BIBentryALTinterwordspacing
A.~Ashkenazy and S.~Zini, ``Attacking machine learning,'' Blogpost: \url{https://skylightcyber.com/2019/07/18/cylance-i-kill-you/Cylance - Adversarial Machine Learning Case Study.pdf}, 2019. [Online]. Available: \url{https://skylightcyber.com/2019/07/18/cylance-i-kill-you/Cylance - Adversarial Machine Learning Case Study.pdf}
\BIBentrySTDinterwordspacing

\bibitem[Joyce et~al.(2021{\natexlab{b}})Joyce, Raff, and Nicholas]{Joyce2021}
R.~J. Joyce, E.~Raff, and C.~Nicholas, ``Rank-1 {Similarity} {Matrix} {Decomposition} {For} {Modeling} {Changes} in {Antivirus} {Consensus} {Through} {Time},'' in \emph{Proceedings of the {Conference} on {Applied} {Machine} {Learning} for {Information} {Security}}, 2021, arXiv: 2201.00757v1.

\bibitem[Pendlebury et~al.(2019)Pendlebury, Pierazzi, Jordaney, Kinder, and Cavallaro]{235493}
\BIBentryALTinterwordspacing
F.~Pendlebury, F.~Pierazzi, R.~Jordaney, J.~Kinder, and L.~Cavallaro, ``{TESSERACT}: {Eliminating} {Experimental} {Bias} in {Malware} {Classification} across {Space} and {Time},'' in \emph{28th {USENIX} {Security} {Symposium} ({USENIX} {Security} 19)}.\hskip 1em plus 0.5em minus 0.4em\relax Santa Clara, CA: USENIX Association, Aug. 2019, pp. 729--746. [Online]. Available: \url{https://www.usenix.org/conference/usenixsecurity19/presentation/pendlebury}
\BIBentrySTDinterwordspacing

\bibitem[Schvartzman et~al.(2024)Schvartzman, Sarussi, Ashkenazi, Tocker, Shohet, et~al.]{schvartzman2024new}
I.~Schvartzman, R.~Sarussi, M.~Ashkenazi, Y.~Tocker, T.~F. Shohet \emph{et~al.}, ``A new dataset and methodology for malicious url classification,'' \emph{The AAAI-25 Workshop on Artificial Intelligence for Cyber Security (AICS)}, 2024.

\bibitem[Kantchelian et~al.(2013)Kantchelian, Afroz, Huang, Islam, Miller, Tschantz, Greenstadt, Joseph, and Tygar]{kantchelian2013approaches}
\BIBentryALTinterwordspacing
A.~Kantchelian, S.~Afroz, L.~Huang, A.~C. Islam, B.~Miller, M.~C. Tschantz, R.~Greenstadt, A.~D. Joseph, and J.~D. Tygar, ``Approaches to adversarial drift,'' in \emph{Proceedings of the 2013 ACM Workshop on Artificial Intelligence and Security}, ser. AISec '13.\hskip 1em plus 0.5em minus 0.4em\relax New York, NY, USA: Association for Computing Machinery, 2013, pp. 99--110. [Online]. Available: \url{https://doi.org/10.1145/2517312.2517320}
\BIBentrySTDinterwordspacing

\bibitem[Joyce et~al.(2025{\natexlab{b}})Joyce, Miller, Roth, Zak, Zaresky-Williams, Anderson, Raff, and Holt]{joyce_ember2024_2025}
\BIBentryALTinterwordspacing
R.~J. Joyce, G.~Miller, P.~Roth, R.~Zak, E.~Zaresky-Williams, H.~Anderson, E.~Raff, and J.~Holt, ``{EMBER2024} - {A} {Benchmark} {Dataset} for {Holistic} {Evaluation} of {Malware} {Classifiers},'' in \emph{Proceedings of the 31st {ACM} {SIGKDD} {Conference} on {Knowledge} {Discovery} and {Data} {Mining} {V}.2}, ser. {KDD} '25.\hskip 1em plus 0.5em minus 0.4em\relax New York, NY, USA: Association for Computing Machinery, Aug. 2025, pp. 5516--5526. [Online]. Available: \url{https://dl.acm.org/doi/10.1145/3711896.3737431}
\BIBentrySTDinterwordspacing

\bibitem[Harang and Rudd(2020)]{Harang2020}
\BIBentryALTinterwordspacing
R.~Harang and E.~M. Rudd, ``{SOREL}-{20M}: {A} {Large} {Scale} {Benchmark} {Dataset} for {Malicious} {PE} {Detection},'' \emph{arXiv}, 2020, arXiv: 2012.07634. [Online]. Available: \url{http://arxiv.org/abs/2012.07634}
\BIBentrySTDinterwordspacing

\bibitem[Krasser et~al.(2025)Krasser, Spurlock, Radu, Moon, Korn, Seth, and Bausewein]{Krasser2025malware}
\BIBentryALTinterwordspacing
S.~Krasser, J.~Spurlock, M.~Radu, B.~Moon, A.~Korn, M.~Seth, and C.~Bausewein, \emph{Machine Learning-Based Malware Detection in a Production Setting}.\hskip 1em plus 0.5em minus 0.4em\relax Cham: Springer Nature Switzerland, 2025, pp. 119--142. [Online]. Available: \url{https://doi.org/10.1007/978-3-031-66245-4_5}
\BIBentrySTDinterwordspacing

\bibitem[Langner(2011)]{Langner2011}
\BIBentryALTinterwordspacing
R.~Langner, ``Stuxnet: {Dissecting} a {Cyberwarfare} {Weapon},'' \emph{IEEE Security \& Privacy Magazine}, vol.~9, no.~3, pp. 49--51, May 2011, iSBN: 1540-7993. [Online]. Available: \url{http://ieeexplore.ieee.org/document/5772960/}
\BIBentrySTDinterwordspacing

\bibitem[Nguyen et~al.(2021)Nguyen, Raff, Nicholas, and Holt]{Nguyen2021}
\BIBentryALTinterwordspacing
A.~T. Nguyen, E.~Raff, C.~Nicholas, and J.~Holt, ``Leveraging {Uncertainty} for {Improved} {Static} {Malware} {Detection} {Under} {Extreme} {False} {Positive} {Constraints},'' in \emph{{IJCAI}-21 1st {International} {Workshop} on {Adaptive} {Cyber} {Defense}}, 2021, arXiv: 2108.04081. [Online]. Available: \url{http://arxiv.org/abs/2108.04081}
\BIBentrySTDinterwordspacing

\bibitem[Barros et~al.(2022)Barros, Chagas, Oliveira, Queiroz, and Ramos]{barros2022malware}
P.~H. Barros, E.~T. Chagas, L.~B. Oliveira, F.~Queiroz, and H.~S. Ramos, ``Malware-smell: A zero-shot learning strategy for detecting zero-day vulnerabilities,'' \emph{Computers \& Security}, vol. 120, p. 102785, 2022.

\bibitem[Akdeniz et~al.(2025)Akdeniz, Ye{\c{s}}ilkaya, K{\"o}se, {\"U}nal, and {\c{S}}en]{akdeniz2025voltron}
M.~T. Akdeniz, Z.~Ye{\c{s}}ilkaya, {\.I}.~K{\"o}se, {\.I}.~U. {\"U}nal, and S.~{\c{S}}en, ``Voltron: Detecting unknown malware using graph-based zero-shot learning,'' \emph{arXiv preprint arXiv:2507.04275}, 2025.

\bibitem[Zeng et~al.(2024)Zeng, Zhou, Lou, Ng, Yau, and Winslett]{zeng2024unleashing}
H.~Zeng, P.~Zhou, X.~Lou, Z.~W. Ng, D.~K. Yau, and M.~Winslett, ``Unleashing the power of unlabeled data: A self-supervised learning framework for cyber attack detection in smart grids,'' \emph{arXiv preprint arXiv:2405.13965}, 2024.

\bibitem[Siddique et~al.(2019)Siddique, Akhtar, Aslam~Khan, and Kim]{siddique_kdd_2019}
\BIBentryALTinterwordspacing
K.~Siddique, Z.~Akhtar, F.~Aslam~Khan, and Y.~Kim, ``{KDD} {Cup} 99 {Data} {Sets}: {A} {Perspective} on the {Role} of {Data} {Sets} in {Network} {Intrusion} {Detection} {Research},'' \emph{Computer}, vol.~52, no.~2, pp. 41--51, Feb. 2019. [Online]. Available: \url{https://ieeexplore.ieee.org/document/8672520}
\BIBentrySTDinterwordspacing

\bibitem[Divekar et~al.(2018)Divekar, Parekh, Savla, Mishra, and Shirole]{divekar_benchmarking_2018}
\BIBentryALTinterwordspacing
A.~Divekar, M.~Parekh, V.~Savla, R.~Mishra, and M.~Shirole, ``Benchmarking datasets for {Anomaly}-based {Network} {Intrusion} {Detection}: {KDD} {CUP} 99 alternatives,'' in \emph{2018 {IEEE} 3rd {International} {Conference} on {Computing}, {Communication} and {Security} ({ICCCS})}, Oct. 2018, pp. 1--8. [Online]. Available: \url{https://ieeexplore.ieee.org/document/8586840}
\BIBentrySTDinterwordspacing

\bibitem[Gelman et~al.(2023)Gelman, Taoufiq, V{\"o}r{\"o}s, and Berlin]{gelman_that_2023}
\BIBentryALTinterwordspacing
B.~Gelman, S.~Taoufiq, T.~V{\"o}r{\"o}s, and K.~Berlin, ``That {Escalated} {Quickly}: {An} {ML} {Framework} for {Alert} {Prioritization},'' Feb. 2023, arXiv:2302.06648 [cs]. [Online]. Available: \url{http://arxiv.org/abs/2302.06648}
\BIBentrySTDinterwordspacing

\bibitem[Raff et~al.(2020{\natexlab{b}})Raff, Filar, and Holt]{Raff2020d}
\BIBentryALTinterwordspacing
E.~Raff, B.~Filar, and J.~Holt, ``Getting {Passive} {Aggressive} {About} {False} {Positives}: {Patching} {Deployed} {Malware} {Detectors},'' in \emph{2020 {International} {Conference} on {Data} {Mining} {Workshops} ({ICDMW})}.\hskip 1em plus 0.5em minus 0.4em\relax IEEE, Nov. 2020, pp. 506--515. [Online]. Available: \url{https://ieeexplore.ieee.org/document/9346444/}
\BIBentrySTDinterwordspacing

\bibitem[Sarker et~al.(2024)Sarker, Janicke, Mohsin, Gill, and Maglaras]{sarker2024explainable}
I.~H. Sarker, H.~Janicke, A.~Mohsin, A.~Gill, and L.~Maglaras, ``Explainable ai for cybersecurity automation, intelligence and trustworthiness in digital twin: Methods, taxonomy, challenges and prospects,'' \emph{ICT express}, vol.~10, no.~4, pp. 935--958, 2024.

\bibitem[Rastogi et~al.(2025)Rastogi, Dhanuka, Saxena, Mairal, and Nguyen]{rastogi2025survey}
N.~Rastogi, D.~Dhanuka, A.~Saxena, P.~Mairal, and L.~Nguyen, ``Survey perspective: The role of explainable ai in threat intelligence,'' \emph{arXiv preprint arXiv:2503.02065}, 2025.

\bibitem[Capuano et~al.(2022)Capuano, Fenza, Loia, and Stanzione]{capuano2022explainable}
N.~Capuano, G.~Fenza, V.~Loia, and C.~Stanzione, ``Explainable artificial intelligence in cybersecurity: A survey,'' \emph{Ieee Access}, vol.~10, pp. 93\,575--93\,600, 2022.

\bibitem[Madamidola et~al.(2025)Madamidola, Ngobigha, and Ez-zizi]{madamidola2025detecting}
O.~A. Madamidola, F.~Ngobigha, and A.~Ez-zizi, ``Detecting new obfuscated malware variants: A lightweight and interpretable machine learning approach,'' \emph{Intelligent Systems with Applications}, vol.~25, p. 200472, 2025.

\bibitem[Paltun et~al.(2025)Paltun, Fuladi, and El~Malki]{paltun2025robust}
B.~G. Paltun, R.~Fuladi, and R.~El~Malki, ``Robust intrusion detection system with explainable artificial intelligence,'' in \emph{2025 Joint European Conference on Networks and Communications \& 6G Summit (EuCNC/6G Summit)}.\hskip 1em plus 0.5em minus 0.4em\relax IEEE, 2025, pp. 145--150.

\bibitem[Mohale and Obagbuwa(2025)]{mohale2025evaluating}
V.~Z. Mohale and I.~C. Obagbuwa, ``Evaluating machine learning-based intrusion detection systems with explainable ai: enhancing transparency and interpretability,'' \emph{Frontiers in Computer Science}, vol.~7, p. 1520741, 2025.

\bibitem[He et~al.(2022)He, Ji, and Huang]{he2022illuminati}
H.~He, Y.~Ji, and H.~H. Huang, ``Illuminati: Towards explaining graph neural networks for cybersecurity analysis,'' in \emph{2022 IEEE 7th European symposium on security and privacy (EuroS\&P)}.\hskip 1em plus 0.5em minus 0.4em\relax IEEE, 2022, pp. 74--89.

\end{thebibliography}

\end{document}